# Mössbauer Spectroscopic Study of Amorphous Fe-Gluconate

Stanisław M. Dubiel[1*], Łukasz Gondek[1], Jan Żukrowski[2]

AGH University of Science and Technology, [1]Faculty of Physics and Applied Computer Science, [2]Academic Center for Materials and Nanotechnology, PL-30-059 Kraków, Poland

## Abstract

Amorphous Fe-gluconate was studied by means of the X-ray diffraction and Mössbauer spectroscopy. Spectra measured in the temperature range between 78 and 295 K were analysed in terms of three doublets using a thin absorber approximation method. Two of the doublets were associated with the major Fe(II) phase (~72%) and one with the minor Fe(III) phase (~28%). Based on the obtained results the following quantities characteristic of lattice dynamical properties were determined: Debye temperature from the temperature dependence of the center shift and that of the spectral area (recoil-free factor), force constant, change of the kinetic and potential energies of vibrations. The lattice vibrations of Fe ions present in both ferrous and ferric phases are not perfectly harmonic, yet on average they are. Similarities and differences to the crystalline Fe-gluconate are also reported.

* Corresponding author: Stanislaw M. Dubiel@fis.agh.edu.pl (S. M. Dubiel)

# 1. Introduction

Ferrous gluconate, a salt of the gluconic acid, has mainly medical and food additive applications. Concerning the former it is effectively used in the treatment of hypochromic anemia and marked under various trade names such as Ascofer®, Fergon®, Ferate®, Ferralet®, FE-40®, Gluconal FE® and Simron®, to list just some of them. Regarding the latter, it is used for coloring foods, e. g. Black olives and beverages and is labelled by the E579 code in Europe. It is worth mentioning that Fe-gluconate was also applied in the metallurgical industry as an effective inhibitor for carbon steel [1], and gluconate-based electrolytes were successfully used to electroplate various metals [2] or alloys [3]. Its chemical formula reads $C_{12}H_{22}FeO_{14}$ (dehydrated) and $C_{12}H_{22}FeO_{14} \cdot 2H_2O$ (hydrated) and iron, whose concentrations lies between 11.8 and 12.5 percent, is present as divalent - $Fe^{2+}$ or Fe(II) – ion which is soluble, hence assimilable by humans. However, a minor fraction (~10-15% relative to the major fraction) of ferric ($Fe^{3+}$) or Fe(III) iron was detected by Mössbauer-effect studies [4-8]. Its origin is unknown and it can be either soluble or insoluble. Clinical studies give evidence that ferric iron medicaments based on iron salts have poorer absorption than the ferrous ones [9], hence they are less effective in the treatment of anemia diseases. The insoluble ferric iron is useless for such treatments, hence its presence in medicaments is undesired. In these circumstances any attempt aimed at the identification of the minor fraction present in the ferrous gluconate is of interest because it can help to get read of it. In a given structure, that can be either crystalline [10] or amorphous [11], the Fe(III) ions should be stronger bounded than the Fe(II) ones. Consequently, their lattice-dynamical properties, such as the value of the Debye temperature, should be different than those of the ferrous ions. The Mössbauer spectroscopy is known to be relevant technique to study the issue. However, our recent study performed in the temperature range of 80-310K on a crystalline form of the Fe-gluconate did not show any significant difference in the lattice-dynamical behavior of the two types of Fe-ions [12]. In order to shed more light on the issue we have carried out similar measurements on an amorphous form of this compound. The results obtained are presented and discussed in this paper.

## 2. Experimental

### 2.1. Sample

The amorphous sample was prepared by dissolving the crystalline specimen in distilled water, then drying it at 50°C in air. The obtained sample's color was dark green, similarly to amorphous specimen obtained in a different route as reported in [11]. The X-ray diffraction (XRD) studies were carried out using Panalytical Empyrean diffractometer with Cu $K_\alpha$ radiation. For high-temperature XRD studies the Anton Paar HTK 1200 N chamber was used.

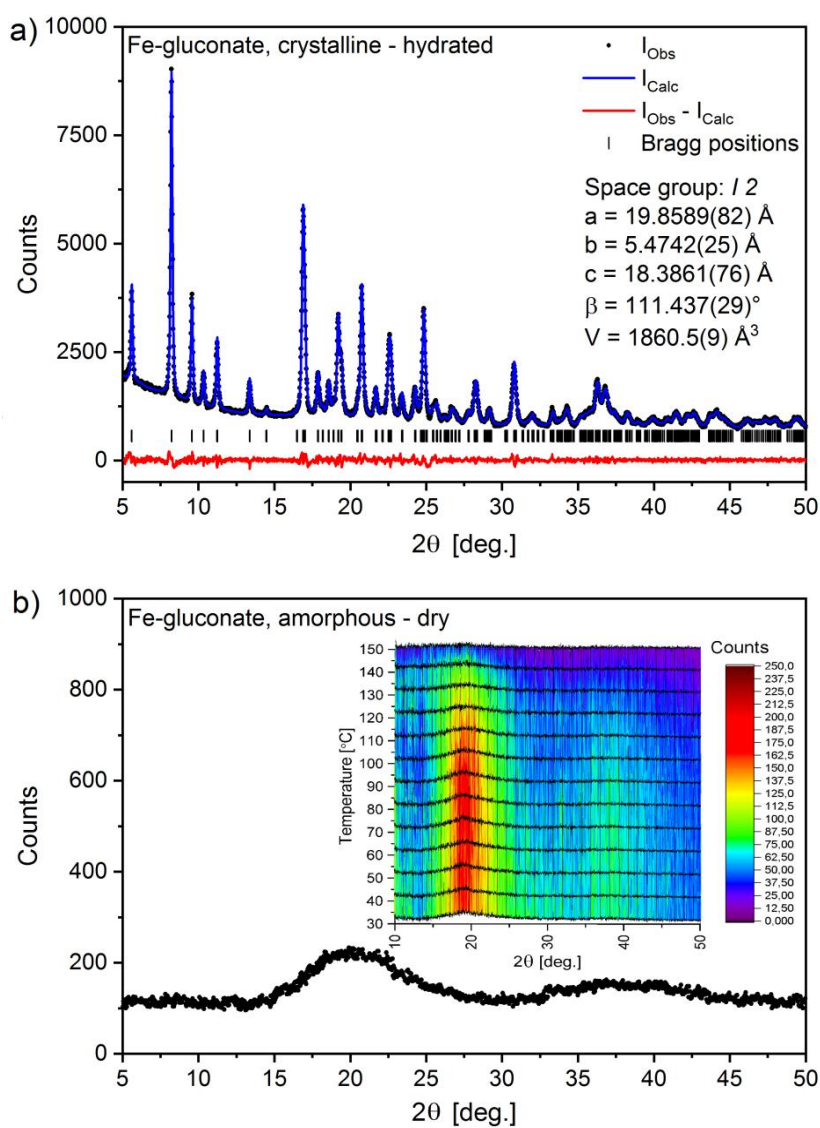

Fig. 1 X-ray diffraction studies of Fe-gluconate: a) crystalline, hydrated specimen and b) amorphous, dried sample with thermal evolution.

The X-ray diffraction (XRD) studies were carried out using Panalytical Empyrean diffractometer with Cu $K_\alpha$ radiation. For high-temperature XRD studies the Anton Paar HTK 1200 N chamber was used.

XRD patterns of crystalline Fe-gluconate and amorphous specimen are compared in Fig. 1. The crystalline (hydrate) Fe-gluconate, which was the parent compound, exhibits excellent structural quality as can be seen in Fig. 1a. The reflections can be entirely indexed by the monoclinic *I*2 (No. 5) space group. The Le Bail refinements let us estimate the lattice parameters as given in the inset to the Fig. 1a. The obtained values are similar to those reported in ref. [10]. It is worth mentioning that unindexed diffraction pattern of Fe-gluconate present in the PDF2 database (entry #00-005-0257) seems to be inaccurate according to [10] and our results.

It is apparent that in amorphous sample there are no signs of the crystalline order, which is characteristic of the parent compound. The amorphous specimen exhibits a very broad maximum at around 20° of 2θ, which corresponds to interplanar distances d ~ 4.6 Å. The additional, much smaller maxima can be noticed at 37° of 2θ, which corresponds to half of the aforementioned distance ~ 2.3 Å. The temperature evolution of the amorphous specimen is presented in Fig. 1b. It was evidenced that thermal decomposition in air takes place at temperatures around 150°C, which is similar to the crystalline Fe-gluconate.

**2.2. Measurements and Analysis**

$^{57}$Fe Mössbauer spectra were recorded in a transmission geometry using a standard spectrometer (Wissel GmbH) with a drive working in a constant acceleration mode. Each measurement was recorded in 1024 channels of a multichannel analyzer. A powdered sample with iron concentration of ~10 mg per cm$^2$ was placed in a Janis SVT-200 cryostat and the temperature of measurements was changed between 78 and 295 K using liquid nitrogen as a coolant. The 14.4 keV gamma rays were provided by a $^{57}$Co/Rh source kept at room temperature. Its activity enabled recording a good quality spectrum within a 3 days run. The temperature of the sample was kept constant within ±0.1 K accuracy during each measurement. Examples of two spectra are shown in Fig. 2. They are similar yet not identical to those recorded on the crystalline compound [7,12]. The most visible difference is a

higher intensity of a minor component (about 2-fold) and more symmetrical outermost lines.

The spectra were analyzed in terms of three doublets using a thin approximation protocol: two of them viz. D1 and D2, based on their spectral parameters, were associated with $Fe^{2+}$ or Fe(II) ions, and one, D3, with $Fe^{3+}$ or Fe(III) ions. The following spectral parameters were fitted: spectral area (Ak), center shift (CSk), quadrupole splitting (QSk), linewidth (Gk) where k=1, 2, 3. This analysis yielded statistically very good fits, and the best-fit parameters are displayed in Table 1.

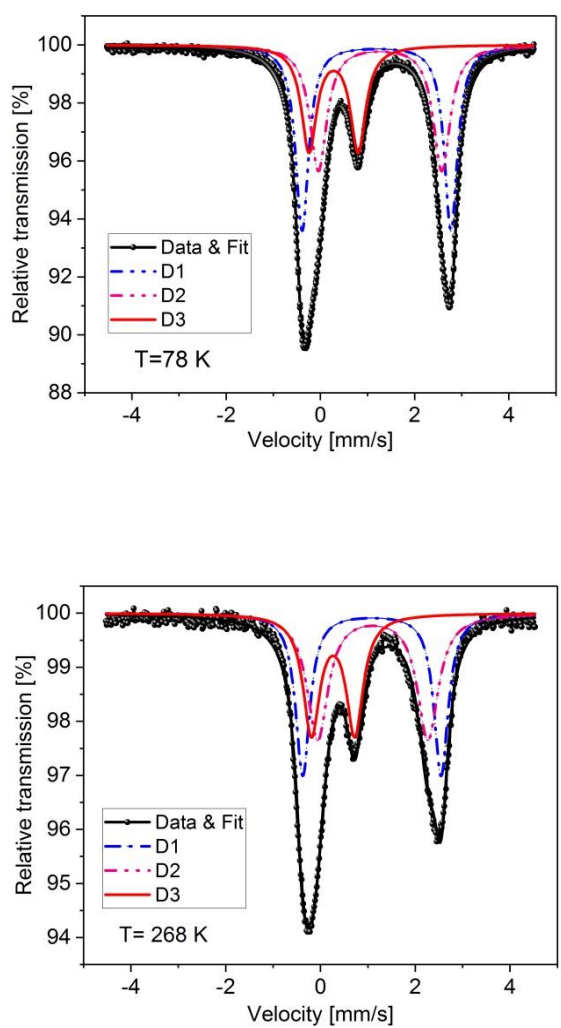

Fig. 2 Examples of two spectra recorded at various temperatures shown. Indicated are three sub spectra D1, D2 and D3 in terms of which the spectra were analyzed.

Table 1

Best-fit spectral parameters obtained by fitting Mössbauer spectra of the amorphous sample of the Fe-gluconate with the thin absorber approximation method. The meaning of the symbols is as follows: $T$ – temperature, Ak – relative spectral area for three components (k=1,2,3), CSk – center shift for the three components, QSk – quadrupole splitting for the three components, Gk – full linewidth at half maximum for the three components, ak – quantity proportional to the recoil-free fraction (normalized spectral area for the $k$-th component). Typical errors: ±0.5% for A1 and A2, ±1% for A3; ±0.003 for CS1 and CS2 and ±0.008 for CS3; ±0.01 for QS1 and QS2 and ± 0.02 for QS3; ±0.01 for G1, G2 and G3.

| T/K | A1 | A2 | A3 | CS1* | CS2* | CS3* | QS1* | QS2* | QS3* | G1* | G2* | G3* | a1 | a2 | a3 |
|---|---|---|---|---|---|---|---|---|---|---|---|---|---|---|---|
| 78 | 36.5 | 36.5 | 27.1 | 1.194 | 1.179 | 0.39 | 3.17 | 2.81 | 0.82 | 0.31 | 0.40 | 0.43 | 1.00 | 1.00 | 1.00 |
| 100 | 34.6 | 37.7 | 28.0 | 1.186 | 1.171 | 0.385 | 3.16 | 2.80 | 0.825 | 0.31 | 0.41 | 0.43 | 0.905 | .96 | .96 |
| 120 | 34.2 | 37.6 | 28.2 | 1.175 | 1.161 | 0.385 | 3.13 | 2.77 | 0.82 | 0.31 | 0.42 | 0.43 | .85 | .93 | .91 |
| 140 | 33.2 | 38.7 | 28.1 | 1.169 | 1.149 | 0.38 | 3.11 | 2.74 | 0.81 | 0.31 | 0.44 | 0.44 | .80 | .92 | .89 |
| 160 | 34.5 | 37.4 | 28.1 | 1.159 | 1.133 | 0.37 | 3.08 | 2.70 | 0.825 | 0.33 | 0.49 | 0.44 | .78 | .86 | .87 |
| 178 | 35.1 | 36.8 | 28.1 | 1.146 | 1.123 | 0.36 | 3.05 | 2.64 | 0.81 | 0.34 | 0.49 | 0.44 | .76 | .77 | .82 |
| 196 | 34.0 | 37.3 | 28.8 | 1.137 | 1.110 | 0.355 | 3.03 | 2.61 | 0.80 | 0.34 | 0.50 | 0.45 | .69 | .76 | .80 |
| 210 | 34.0 | 37.4 | 28.6 | 1.127 | 1.091 | 0.35 | 3.01 | 2.58 | 0.81 | 0.35 | 0.49 | 0.46 | .66 | .73 | .76 |
| 224 | 33.9 | 37.7 | 28.5 | 1.116 | 1.077 | 0.34 | 2.96 | 2.53 | 0.82 | 0.38 | 0.53 | 0.47 | .64 | .66 | .74 |
| 238 | 34.7 | 36.1 | 29.2 | 1.108 | 1.065 | 0.335 | 2.96 | 2.52 | 0.805 | 0.37 | 0.53 | 0.46 | .635 | .65 | .71 |
| 252 | 35.0 | 36.2 | 28.8 | 1.101 | 1.056 | 0.33 | 2.93 | 2.46 | 0.80 | 0.37 | 0.55 | 0.45 | .61 | .62 | .67 |
| 268 | 35.0 | 37.6 | 28.4 | 1.091 | 1.042 | 0.32 | 2.90 | 2.44 | 0.78 | 0.39 | 0.56 | 0.47 | .57 | .53 | .65 |
| 295 | 34.9 | 36.9 | 28.2 | 1.077 | 1.025 | 0.315 | 2.84 | 2.40 | 0.77 | 0.38 | 0.56 | 0.46 | .49 | .46 | .52 |

* in mm/s. Values of center shifts are relative to the $^{57}$Co/Rh source at RT.

## 3. Results and Discussion

### 3. 1. Debye Temperature

The Debye temperature, $T_D$, regarded as measure of a lattice stiffness, can be determined either from a temperature dependence of (1) center shift, CS, or from (2) recoil-free fraction, $f$. The former can be expressed as follows:

$$CS(T) = IS(T) + SOD(T) \qquad (1)$$

Where IS stays for the isomer shift and SOD is the so-called second order Doppler shift i.e. a quantity related to a non-zero mean value of the square velocity of vibrations, <v²>, hence kinetic energy. Assuming that the phonon spectrum can be described by the Debye model, and taking into account that *IS* hardly depends on temperature, hence it can be neglected [13,14], the temperature dependence of *CS* can be thus related to $T_D$ via the second term in eq. (1) that has the following form [14]:

$$CS(T) = IS(0) - \frac{3k_B T}{2mc}\left(\frac{3T_D}{8T} + 3\left(\frac{T}{T_D}\right)^3 \int_0^{T_D/T} \frac{x^3}{e^x - 1}dx\right) \quad (2)$$

Here *m* stays for the mass of the Fe atom, $k_B$ is the Boltzmann constant, *c* is the speed of light, and $x = \hbar\omega/kT$ ($\omega$ being frequency of vibrations).

An example of a *CS1(T)* dependence is presented in Fig. 3, and all $\Theta_D$-values obtained for the three components using this approach are displayed in Table 3.

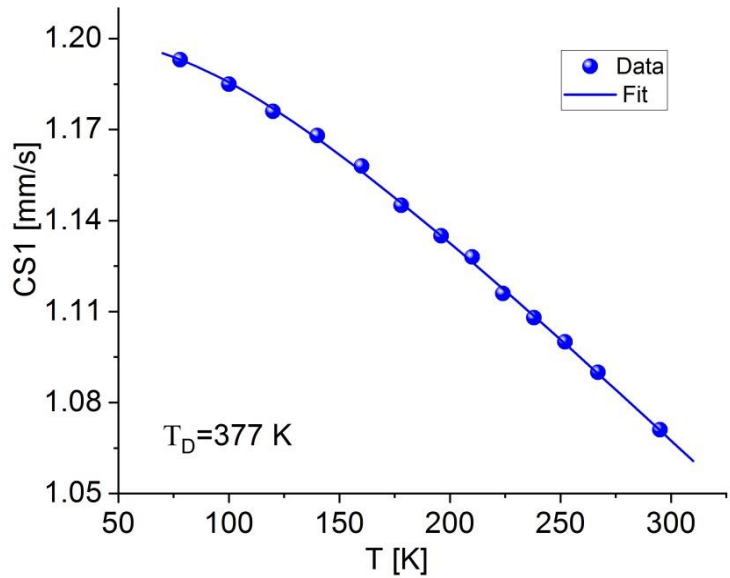

Fig. 3 Temperature dependence of the center shift of the component D1, *CS1*. The solid line represents the best fit of the data to eq. (2).

Table 2.

Values of the Debye temperature, $T_D$, as determined from the temperature dependence of the center shift, $CS_k$, of particular components (k=1, 2, 3), and that of the average center shift, $<CS>$, as well as from the quantity proportional to the recoil-free fraction, $a_k$. For comparison, corresponding values determined for the crystalline sample (Ref. 12) have been added.

| | Amorphous sample | | | | | | |
|---|---|---|---|---|---|---|---|
| | CS1 | CS2 | CS3 | <CS> | a1 | a2 | a3 |
| $T_D$[K] | 377(9) | 231(12) | 672(29) | 276(24) | 204(4) | 209(4) | 241(9) |
| | Crystalline sample [12] | | | | | | |
| | CS12 | | CS3 | <CS> | a12 | | a3 |
| $T_D$[K] | 437(21) | | 346(149) | 423(40) | 206(12) | | 226(52) |

As can be seen the $T_D$-values are characteristic of the sub spectrum. In particular, $T_D$ determined from the CS1(T) is by ~50% biger than the one found from the CS2(T). This finding testifies to a heterogenous or, at least, distorted structure of the major component of the Fe-gluconate. $Fe^{2+}$ ions associated with the D2 component have lower values of QS and very similar CS ones. This means that they occupy positions with a slightly higher or less deformed symmetry than those associated with the D1 subspectrum. Unfortunately, lack of information on structural positions of Fe atoms in the Fe-gluconate does not permit to interpret the data in a more quantitative way. Significantly higher value was determined from the CS3(T) dependence, hence depicting the $Fe^{3+}$ ions. This can have two reasons: (1) a stronger binding of $Fe^{3+}$ ions due to their larger charge and/or (2) different crystallographic structure of the minor component of the studied compound. Our recent XRD studies give evidence that the crystallographic structure of Fe(II) and Fe(III) ions is the same [15]. $T_D$ can alternatively be figured out from a temperature dependence of the recoil-free fraction, $f=exp(-k^2<x^2>)$, k being the wave vector of the gamma rays and $<x^2>$ stands for the mean square amplitude of vibrations. In the frame of the Debye model the $f$-$T_D$ relationship reads as follows [16]:

$$f = \exp\left[\frac{-6E_R}{k_B T_D}\left\{\frac{1}{4} + \left(\frac{T}{T_D}\right)^2 \int_0^{T_D/T} \frac{x\, dx}{e^x - 1}\right\}\right] \quad (3)$$

Where $E_R$ is the recoil kinetic energy, $k_B$ is Boltzmann constant.

In the thin absorber approximation $f$ is proportional to a spectral area, $A$, so the latter is used in a practical application of eq.(3) in order to determine $T_D$. However, the value of the spectral area for a given sample and measurements conditions depends on the number of counts. In order to take this into account, we considered a normalized value of $A_k$, $a_k=(A_k/B)/a_k(78K)$, where $B$ stays for the number of counts in background (base line) of a spectrum.

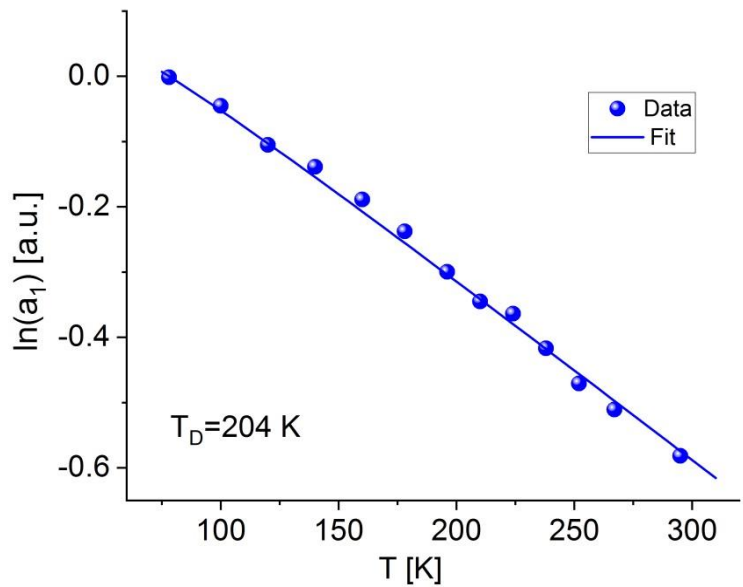

Fig. 4 Temperature dependence of $ln(a_1)$. The solid line stays for the best-fit of the data to eq. (3).

The values of $T_D$ achieved from this spectral parameter for the three components are also displayed in Table 2. Their values are smaller than those found from the $CS(T)$ dependences, yet the $T_D$-value associated with the D3-component is greater than the ones related to D1 and D2 doublets. It is of interest to compare these results with the ones received for a crystalline sample of the Fe-gluconate [12]. Noteworthy, the $T_D$-values derived from the spectral area are practically the same and equal to ~200 K. However, the values acquired from $CS(T)$ are different. Whereas for the crystalline sample they had, within experimental error, similar values (~400 K), for the amorphous sample $T_D$ for the minor phase ($Fe^{3+}$) is significantly larger. In addition, the two components, into which was analyzed the major phase ($Fe^{2+}$) of the amorphous sample, have significantly different values of $T_D$, while this was not so in the case of the crystalline sample. This relationship has been reflected in the values of $T_D$ derived from the temperature dependence of the average center shift, $<CS>(T)$.

### 3. 2. Energetics of Vibrations

It is of interest to express the vibrations of Fe atoms associated with the three components in terms of the underlying kinetic, $E_K$, and potential, $E_P$, energies. The average kinetic, $E_K=0.5m<v^2>$, and potential, $E_P=0.5F<x^2>$ (in harmonic approximation) energies of the lattice vibrations can be determined assuming the $SOD$ and the $f$-factor are known. The force constant is denoted by $F$, $m$ is the mass of an vibrating atom (here $^{57}$Fe) and $c$ is the velocity of light. Taking into account that by definition $SOD=-0.5E_\gamma<v^2>/c^2$, $E_\gamma$ being the energy of the gamma-rays (14.4 keV in the present case), the average kinetic energy can be expressed as follows:

$$E_K = -mc^2 \frac{SOD}{E_\gamma} \qquad (4)$$

The relationship between $E_P$ and $f$ is, in turn, given by the following term:

$$E_P = -\frac{1}{2} F \left(\frac{\hbar c}{E_\gamma}\right)^2 \ln f \qquad (5)$$

$E_K$ can be readily calculated from eq. (4) using the *SOD*-values measured in the Mössbauer experiment, whereas $E_P$ cannot as the value of $F$ has also to be known. Concerning the former the calculated changes of $E_k$ based on formula (4) are plotted vs. temperature in Fig. 5.

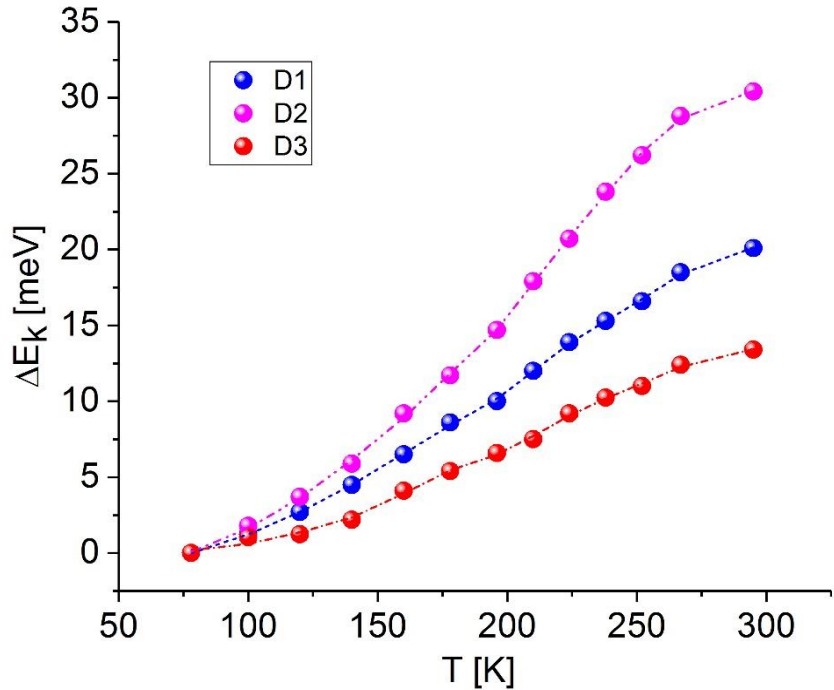

Fig. 5 Change of the kinetic energy, $\Delta E_k = E_k(T) - E_k(78K)$, vs. temperature, $T$, for the three components D1, D2 and D3.

It is worth to observe that the change of the kinetic energy, $\Delta E_k = E_k(T) - E_k(78K)$, significantly depends on the component, being the largest for D2 and the smallest for D3. This gives a strong evidence that the kinetic energy of the lattice vibrations (1) in the ferric phase are unquestionably different than those in the ferrous phase and (2) the ferrous phase is heterogeneous as far as the kinetic energy of lattice vibrations are concerned. Noteworthy, the average change of the kinetic energy, $<\Delta E_k>$=20.1 meV, agrees quite well with the change due to the increase of temperature, $\Delta E=k_B \cdot \Delta T$=18.7 meV what supports our present calculations depicting the energy of vibrations.

Before we will discuss the issue more deeply, it is reasonable to first determine corresponding changes in $E_p$. To this end, as already mentioned, the knowledge of $F$ is necessary. As shown elsewhere [12], this knowledge can be obtained based on a linear correlation between a change of $<v^2>$, $\Delta<v^2>=<v^2>(T)-<v^2>(78K)$, and that of $<x^2>$, $\Delta<x^2>=<x^2>(T)-<x^2>(78K)$. Figure 6 displays such correlations for D1, D2 and D3.

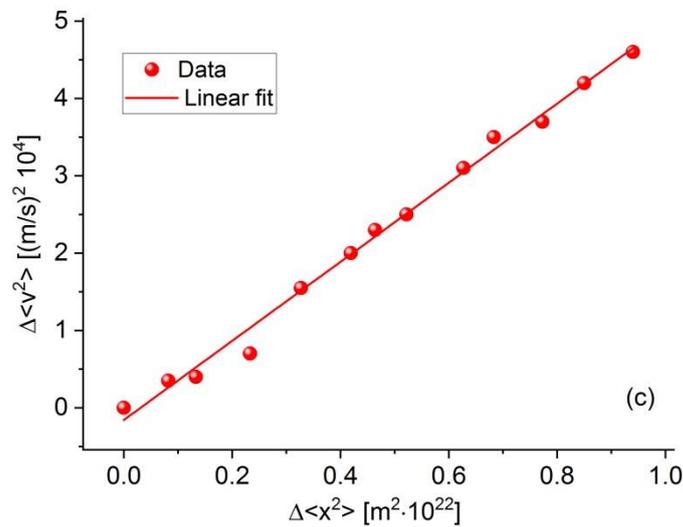

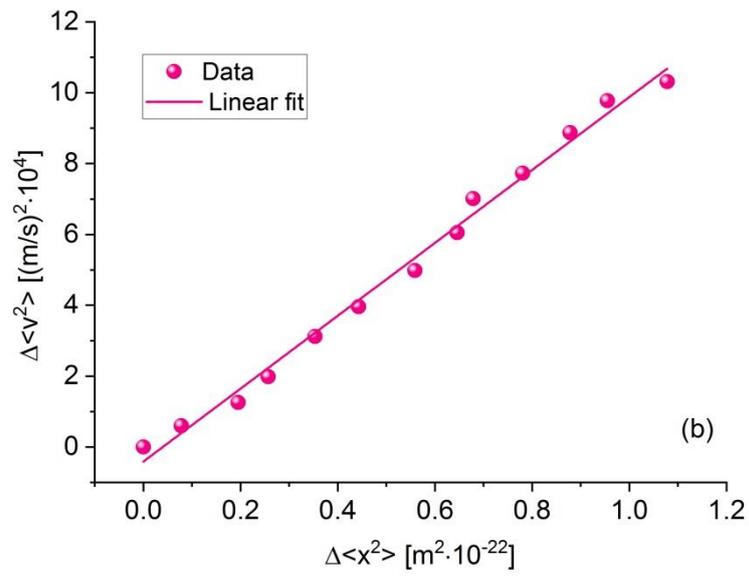

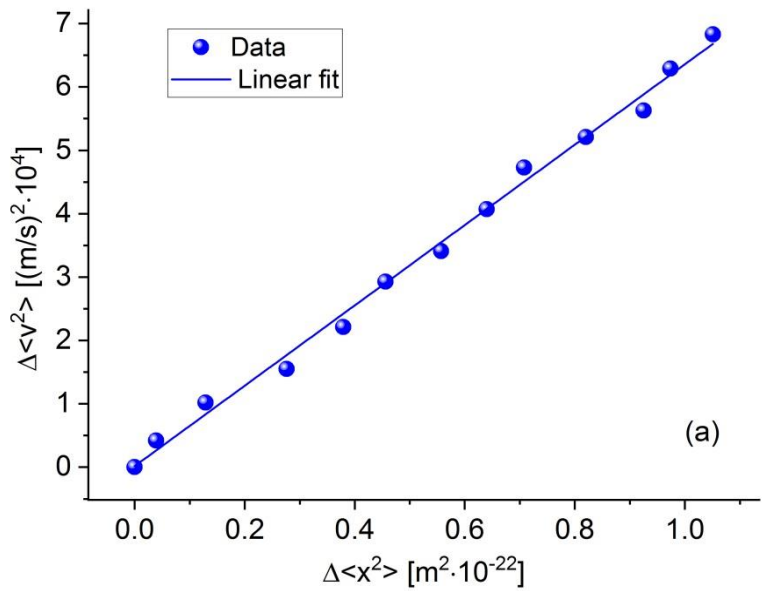

Fig. 6 Relationships between $\Delta\langle v^2\rangle$ and $\Delta\langle x^2\rangle$ for the three subspectra: (a) D1, (b) D2 and (c) D3. The data were fitted to a linear equation. The best fits are marked by solid lines.

The value of $F_k = m \cdot \alpha_k$, where $\alpha_k$ stays for the slope of the line for the $Dk$ component. In this way the following values of the force constant were obtained: $F_1$=50.8 N/m, $F_2$=69.4 N/m and $F_3$=43.6 N/m. As can be seen they are characteristic of the sub spectrum. Interestingly, the value of $F$ as determined for the ferrous Fe-ions in the crystalline sample was equal to 44 N/m [12], hence significantly less than $F_1$ and $F_2$ in the present case. Worth to note, the value of $F_3$ is lower than both $F_1$ and $F_2$ which likely can be understand in terms of a different environment of $Fe^{3+}$ ions than that of the $Fe^{2+}$ ones.

Knowing the $F_k$-values and using the formula (5) we have calculated relative changes of $E_p$, $\Delta E_p(T) = E_p(T) - E_p(78K)$, for D1, D2 and D3. Figure 7 illustrates the obtained results.

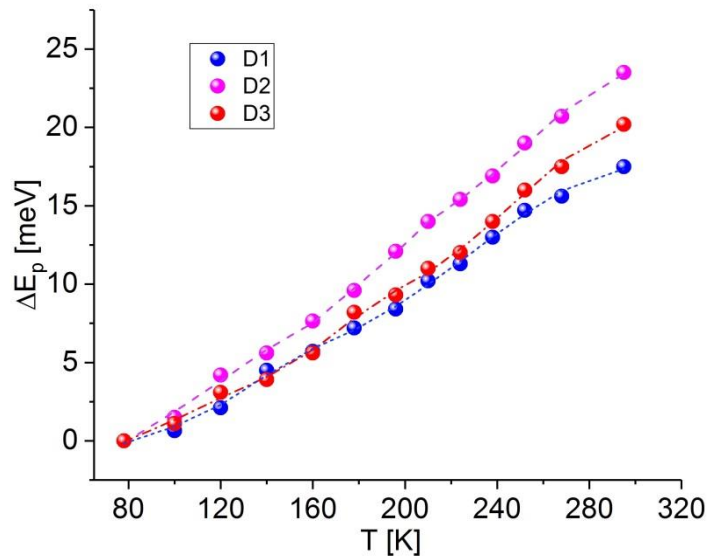

Fig. 7 Change of the potential energy, $\Delta E_p = E_p(T) - E_p(78K)$, vs. temperature, $T$, for the three components D1, D2 and D3.

Here the differences between the three components are small, especially at lower temperatures. In the case of harmonic oscillations the changes of the two forms of the mechanical energy i.e $E_k$ and $E_p$ should be exactly the same. In order to see whether or not this is the case here, we have plotted a relationship between $\Delta E_k$ and $\Delta E_p$ for each component. The results obtained are displayed in Fig.8.

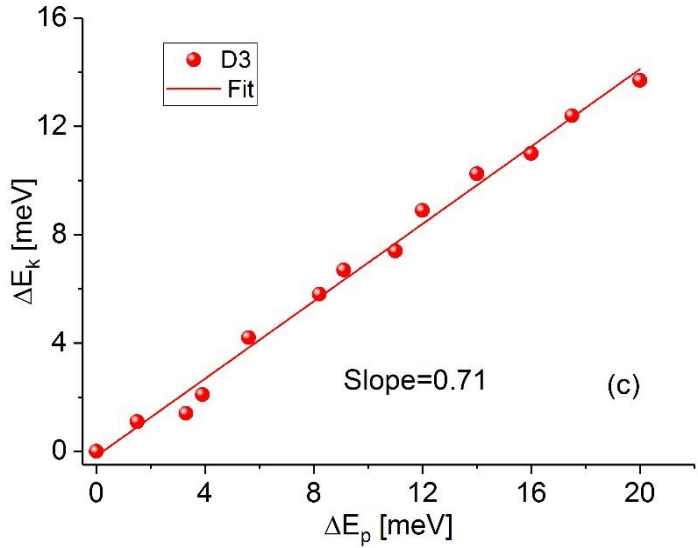

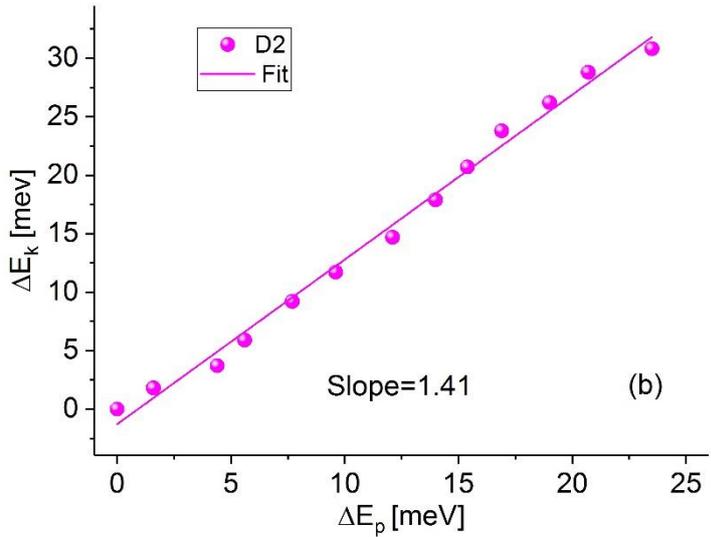

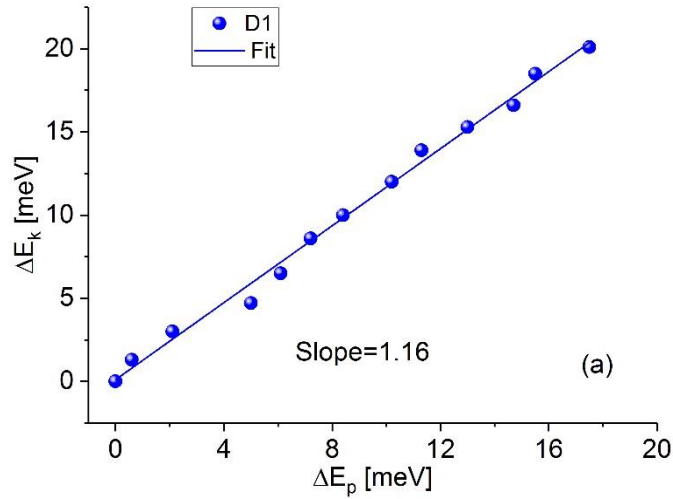

Fig. 8 Relationship between $\Delta E_k$ and $\Delta E_p$ for the three sub spectra D1, D2 and D3. The lines represent the best linear fits to the data.

It follows from Fig. 8 that in each case the slope is different than 1 what means that the vibrations of Fe ions are not strictly harmonic. The least deviation from the harmonicity is found for the D1 sub spectrum and the highest deviation is observed for the D2 component. This difference shows again that the ferrous phase is heterogeneous. An indication of the latter is also seen in a different line width of D1 and D2 components – see Table 1. The Fe ions in the ferric phase also exhibit 30% deviation from the harmonic mode, but here the slope is < 1 what means that the change in the potential energy is larger than the one in the kinetic energy. However, the maximum value of the potential energy change, $<E_p>$=20.4 meV, what perfectly agrees with the corresponding change of the kinetic energy, $<E_k>$=20.1 meV. This means that, on average, the vibrations of Fe-ions present in the studied sample are harmonic.

Concerning the ferrous ions, at least two components, D1 and D2, can be distinguished that not only have small, yet measurable, differences in spectral parameters, but they also differ in the values of the Debye temperature, kinetic energy of vibrations as well as the force constants. Smaller values of *QS2* than those of *QS1* one can indicate that $Fe^{2+}$ ions associated with D2 have a slightly higher symmetry or a less deformed environment. The reduction of the quadrupole splitting of high-spin ferrous ions could be also due to a charge flow. In fact, there is some small difference in the corresponding values of *CS1* and *CS2*. Perhaps the investigated Fe-gluconate exists in form of small particles and the D2 component is associated with $Fe^{2+}$ occupying particles' core while the D1 one with those ferrous ions that are located on the particles' surface or close to it. This issue cannot be answered fully with the present study.

### 3.3. Quadrupole Splitting

It was reported that a temperature dependence of the quadrupole splitting for the ferrous and the ferric ions was different [17]. For the former *QS* significantly decreases with *T* while for the latter the dependence is weak if any, so it can be used to make a distinction between the two forms of high-spin Fe ions. It is thus of interest to verify whether or not this observation is valid in the present case. The temperature dependence of the quadrupole splitting, *QS(T)*, can be satisfactory described by the following phenomenological equation [18,19]:

$$QS(T) = QS(T_o)\left[1 - aT^{3/2}\right] \qquad (4)$$

Figure 9 gives evidence that also presently found *QS(T)* – values can be very well described by this equation.

The T-dependence of QS3 i.e. the one associated with the ferric ($Fe^{3+}$) component is in fact significantly weaker that the corresponding dependences found for the ferrous ($Fe^{2+}$) components. This finding can be regarded as a strong argument that the Fe-ions associated with the minor phase are trivalent.

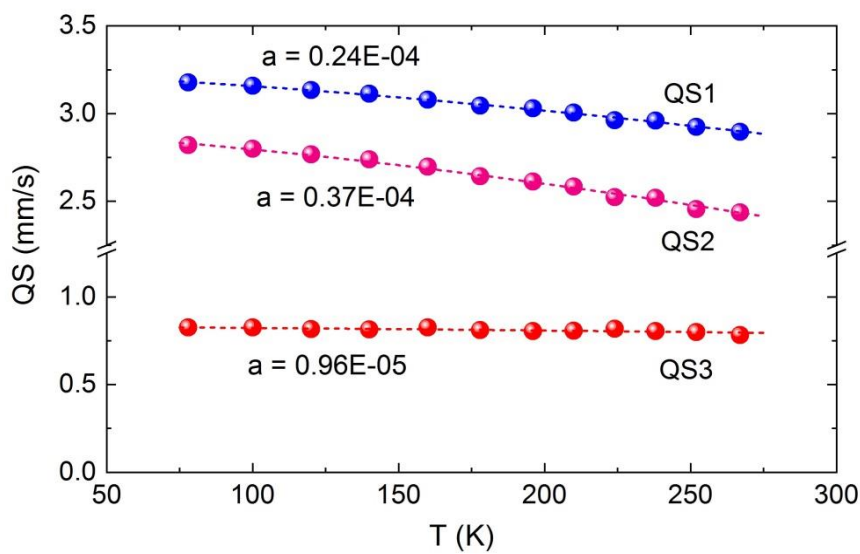

Fig. 9 Temperature dependences of $QSk(T)$ for the three sub spectra (k= 1, 2, 3). Solid lines represent the best fits of the data to eq. (4).

## 4. Conclusions

The presently found results on the amorphous Fe-gluconate permit drawing the following conclusions:

1. Iron atoms are present in form of $Fe^{2+}$ ions (~72%) that can be associated with distorted octahedral environments and $Fe^{3+}$ ones (~28%) that can be related to a distorted tetrahedral environments.

2. The major phase ($Fe^{2+}$) is not homogenous and it can be decomposed, at least, into two equally-shared components having different values of the quadrupole splitting, Debye temperature, force constant, and kinetic energy of vibrations.

3. $Fe^{3+}$ ions have significantly different lattice dynamical properties than the $Fe^{2+}$ ions: much higher value of the Debye temperature determined from the center shift, lower value of the force constant as well as that of the kinetic energy of vibrations.

4. Vibrations of Fe ions associated with the three components deviate slightly from the harmonic ones, yet on average the vibrations are harmonic.

5. Temperature dependence of *QS3* is much weaker than that of *QS1* and *QS2*.

**Acknowledgements**

This work was financed by the Faculty of Physics and Applied Computer Science AGH UST and ACMIN AGH UST statutory tasks within subsidy of Ministry of Science and Higher Education, Warszawa.